\documentclass{article}
\usepackage{graphicx}
\usepackage{amsmath}

\begin{document}

\begin{center}
\vskip3.7 cm

{\LARGE Brief comments on Jackiw-Teitelboim gravity}

\vskip0.3 cm

{\LARGE coupled to Liouville theory}

\vskip1.0 cm

{\large Gast\'{o}n E. Giribet}

\vskip1.0 cm

Instituto de Astronom\'{\i}a y F\'{\i}sica del Espacio, IAFE

C.C. 67 - Suc. 28, (1428) Buenos Aires, Argentina

and

Physics Department, University of Buenos Aires, FCEN-UBA

\vskip2.0 cm

\textbf{Abstract}
\end{center}

Jackiw-Teitelboim gravity with non-vanishing cosmological constant
coupled to Liouville theory is considered as a non-critical string on $d$
dimensional flat spacetime. It is discussed how the presence of
cosmological constant yields additional constraints on the
parameter space of the theory, even when the conformal anomaly is independent of
the cosmological constant. Such constraints agree with the necessary conditions for the tachyon field to be a primary --prelogarithmic--
operator of the worldsheet conformal field theory. Thus,
the linearized tachyon field equation allows to impose the diagonal
condition for the interaction term. We analyze the neutralization of the Liouville mode induced by the coupling
to the Jackiw-Teitelboim Lagrangian. The free field prescription
leads to obtain explicit expressions for three-point correlation functions for the case
of vanishing cosmological constant in terms of a product of Shapiro-Virasoro
integrals. This is a consequence of the mentioned neutralization effect.

\newpage

\vskip0.3 cm

\section{Introduction}

As early observed by Tseytlin, one of the principal motivations to study the dependence of the effective action on the tachyon
field comes from the connection with two-dimensional gravity 
\cite{tseytlin}. In this note, we study the linearized tachyon
field equation within the context of the Jackiw-Teitelboim model of 2D gravity \cite{teitelboim} thought of as a non-linear $\sigma 
$-model describing the worldsheet dynamics of a non-critical string.

This class of analogies existing between two-dimensional models of
gravitation and non-critical string theories was previously discussed in the
literature. For instance, in reference \cite{chamseddine}, the action for
two-dimensional gravity coupled to matter fields was quantized using
conformal field theory (CFT) methods. It was
argued there that the quantization could be carried out without restricting the
dimension of the matter content. To do this, the author of \cite{chamseddine}
considered the case of vanishing cosmological constant $\Lambda =0$, arguing
that such case suffices to illustrate the general features of
the model because, after all, the conformal anomaly results independent of the value of $\Lambda $. In this note, we revise such affirmation. We will show how
additional restrictions actually appear whenever a non-vanishing cosmological
constant $\Lambda $ is considered in the Jackiw-Teitelboim two dimensional
theory of gravitation. The remarkable point is that this is true even if the central charge of the model actually
does not depend of the explicit value of $\Lambda $. Our analysis is carried out by
interpreting the model as a consistent non-critical string theory. Such stringy interpretation is not strictly necessary, but it will allow us to make use of standard 2D CFT techniques to solve the problem.  

In reference \cite{mazzitelli}, the Jackiw-Teitelboim theory was also
interpreted as a noncritical string theory in presence of background fields.
Following this representation, we show that the additional restrictions on
the coupling constant and the parameters of the model arise by solving the
linearized tachyon field equation that come from the interpretation
of the model as a non-critical string $\sigma $-model. The central point is based on the fact that
Jackiw-Teitelboim gravity coupled to Liouville theory and matter fields, as
mentioned, could be geometrically interpreted as a CFT
describing the dynamic of a string on a $d$-dimensional Minkowski target
spacetime with a dilaton field. And, in the case of non-vanishing cosmological constant, the
two-dimensional worldsheet theory consists of the sum of two parts: The first part is a $d$-dimensional free field bosonic theory; the second is an
interaction term, like a screening charge, which in the general case a logarithmic operator. Then, an additional constraint on the
parameters of the theory is required in order to impose the necessary
conditions to turn the interaction term into an integral of a primary operator.
This constraint, which is absent if $\Lambda =0$ because it is automaticaly
satisfied in such case, can be translated into a restriction on the
functional form and value of the central charge of the theory.\footnote{The coupling between two-dimensional
models
of gravity and Liouville theory was also studied in references \cite
{odintsov,mann}.} 

In the following sections, we will study the model in the language of two-dimensional CFT. We will analyze the associated non-linear $%
\sigma $-model which can be interpreted as the non-critical string theory on
the $d$-dimensional flat spacetime coupled to linear dilaton background.
After doing that, we will explicitly compute the three-point correlation
functions in the model, in the case of vanishing cosmological constant. We will show how such observables can be written
in terms of Virasoro-Shapiro integrals due to a neutralization of
the Liouville mode produced by the constant curvature condition. The computation
of the correlation functions will be done using the Coulomb gas prescription.

\section{Jackiw-Teitelboim gravity coupled to Liouville theory}

Let us begin by considering the Jackiw-Teitelboim gravitational action
coupled to Liouville theory and $d-2$ scalar fields representing the matter,
namely 
\begin{equation}
S_{eff}=S_{JT}+S_{L}+S_{M}+S_{ghost}  \label{sjt}
\end{equation}
with
\begin{eqnarray*}
S_{JT} &=&\frac{b}{\pi }\int_{M}d^{2}x\sqrt{\hat{h}}\varphi \left( \tilde{R}+%
\tilde{\Lambda}\right) +\frac{\lambda }{4\pi }\int_{M}d^{2}x\sqrt{\hat{h}}%
\tilde{R} \\
S_{L} &=&\frac{1}{\pi }\int_{M}d^{2}x\sqrt{h}\left( ah^{\alpha \beta
}\partial _{\alpha }\sigma \partial _{\beta }\sigma +QR\sigma +\frac{\mu }{2}%
e^{\gamma \sigma }\right) \\
S_{M} &=&\frac{1}{4\pi }\int_{M}d^{2}x\sqrt{h}h^{\alpha \beta }\partial
_{\alpha }X^{i}\partial _{\beta }X^{j}\delta _{ij}
\end{eqnarray*}
where $a$ and $b$ are (positive) real parameters interpreted as coupling
constant of the Jackiw-Teitelboim Lagrangian and the Liouville Lagrangian
respectively. $\tilde{\Lambda}$ represents the target space cosmological
constant, while $\mu $ is the quoted Liouville cosmological constant. $R$ refers to the Ricci scalar corresponding to the two-dimensional
metric $h_{\alpha \beta }$ on the two-manifold $M$, while $\tilde{R}$
corresponds to the Ricci scalar constructed from the rescaled metric $\hat{h}%
_{\alpha \beta }$ \cite{david}, defined by 
\begin{equation}
\hat{h}_{\alpha \beta }=e^{\gamma \sigma }h_{\alpha \beta }  \label{gauge}
\end{equation}
where $(\alpha ,\beta )\in \{0,1\}.$

The field $\varphi $ is the Jackiw-Teitelboim scalar field, while $\sigma $
is the Liouville scalar field; there exist other $d-2$ scalar fields which
represent the matter, being denoted by $X^{i},$ $i\in \{2,3...d-1\}$. The
Hilbert term also appears in the action (\ref{sjt}) contributing with the
Euler characteristic number $\chi (h)$. $\lambda $ is a real
coupling constant which can be related with the dilaton zero-mode. On the
other hand, the ghost action $S_{ghost}$ needs to be included; it can be
expressed in terms of the $b$-$c$ system as 
\begin{equation}
S_{ghost}=\frac{1}{2\pi }\int_{M}d^{2}x\sqrt{h}h^{\alpha \beta }c^{\rho
}\nabla _{\alpha }b_{\beta \rho }
\end{equation}
Notice that the Jackiw-Teitelboim field $\varphi $ acts as a Lagrange
multiplier, which can be integrated out introducing the constant curvature
constraint 
\begin{equation*}
\tilde{R}+\tilde{\Lambda}=0.
\end{equation*}

As said before, this class of two-dimensional gravitational action
coupled to Liouville theory was studied in different contexts. Indeed, an
interesting analysis of its perturbative aspects was performed in \cite
{mazzitelli,mazzitelli2}. In reference \cite{chamseddine} this action was
quantized by using the path integral approach and conformal field methods
too. It was also studied in the context of $W_{3}$ gravity \cite{mohammedi}. A careful analysis of its classical and quantum features was presented
in \cite{burwick}. In \cite{chamseddine,chamseddine2} it was shown that the
Teitelboim-Jackiw Lagrangian appears as the metric version of a topological
construction based on the $SO(1,2)$ group. This realization relies on treating the two-dimensional models of gravity as a kind of
projective reduction of three-dimensional Chern-Simon theory. See also \cite
{shapiro} for an interesting discussion.

Here, we shall consider the classical model defined by action (\ref{sjt}) as
the starting point. With this purpose, let us define the following variables 
\begin{eqnarray}
\sigma &=&\frac{\kappa }{\gamma }\left( X^{1}+X^{0}\right)  \label{var1} \\
\varphi &=&\frac{\kappa }{\gamma }\left( X^{1}+\xi X^{0}\right)  \label{var2}
\end{eqnarray}
where 
\begin{equation}
\kappa =\frac{\gamma }{2\sqrt{a+\gamma b}} \ , \ \ \xi =-1-\frac{2a}{\gamma b}. 
\end{equation}
These new variables are well defined in terms of the original
coordinates $\varphi $ and $\sigma $ if and only if $a+b\gamma \neq 0$. In terms of the
new coordinates, the above action can be written as follows 
\begin{equation}
S_{eff}=S_{0}+S_{I}+\lambda \chi (h)  \label{sx}
\end{equation}
with
\begin{eqnarray}
S_{0} &=&\frac{1}{4\pi }\int d^{2}z\left( \eta _{\mu \nu }\partial X^{\mu }%
\bar{\partial}X^{\nu }+RQ_{0}X^{0}+RQ_{1}X^{1}\right) ,  \label{sx1} \\
S_{I} &=&\frac{1}{4\pi }\int d^{2}z\left( \xi \Lambda X^{0}+\Lambda
X^{1}+2\mu \right) e^{\kappa (X^{0}+X^{1})}  . \label{interaction}
\end{eqnarray}
Here, $\eta _{\mu \nu }$ denotes the $d$-dimensional Minkowski metric and
the following notation was introduced 
\begin{eqnarray*}
Q_{0} &=&2\frac{Q-b-2a/\gamma }{\sqrt{a+\gamma b}} \\
Q_{1} &=&2\frac{Q+b}{\sqrt{a+\gamma b}}
\end{eqnarray*}
In (\ref{sx1}) and (\ref{interaction}) $z$ and $\bar{z}$ denote the usual
worldsheet complex coordinates, being $\partial =\frac{\partial }{\partial z%
}$ and $\bar{\partial}=\frac{\partial }{\partial \bar{z}}$. Notice that the
cosmological constant $\Lambda $ was also rescaled by a constant factor,
namely $\Lambda =\frac{2b}{\sqrt{a+\gamma b}}\tilde{\Lambda}$. We are
considering Greek indices which run over the $d$-dimensional target spacetime, $(\mu
,\nu )\in \{0,1,...d-1\}$.

It is also important to notice that the constant curvature
constraint can be derived by combining the Euler-Lagrange equation for $%
X^{0} $ and $X^{1}$ fields, obtaining 
\begin{equation*}
e^{-\gamma \sigma }\left( R-\gamma \partial ^{2}\sigma \right) +\tilde{%
\Lambda}=0,
\end{equation*}
which can be written in the usual form $\tilde{R}+\tilde{\Lambda}=0$ if the
relation (\ref{gauge}) is taken into account.

Thus, we have written the two-dimensional action (\ref{sjt}) in terms of a
theory of self-interacting scalar fields coupled to classical gravity in the
conformal gauge (\ref{gauge}); it will be a useful picture to work out the
features of the model.

\section{The conformal field theory}

The above free action $S_{0}$ defines a two-dimensional quantum
CFT, which could be analyzed by the usual techniques of 2D CFT.
Indeed, the holomorphic component of the stress-tensor corresponding to this
model is given by 
\begin{equation}
T(z)=-\frac{1}{2}\eta _{\mu \nu }\partial X^{\mu }\partial X^{\nu }+\frac{%
Q_{0}}{2}\partial ^{2}X^{0}+\frac{Q_{1}}{2}\partial ^{2}X^{1}  \label{t}
\end{equation}
with free field bosonic correlators, 
\begin{equation}
\left\langle X^{\mu }(z)X^{\nu }(w)\right\rangle =-\eta ^{\mu \nu }\ln (z-w)
\label{correlators}
\end{equation}
Thus, it is easy to show that the following operator product expansion holds 
\begin{equation*}
T(z)T(w)=\frac{\frac{d}{2}+\frac{3}{2}(Q_{1}^{2}-Q_{0}^{2})}{(z-w)^{4}}+%
\frac{T(w)}{(z-w)^{2}}+\frac{\partial T(w)}{(z-w)}+...
\end{equation*}
where the ellipses stand for regular terms. From this expansion, we obtain that the central charge of the model is given
by 
\begin{equation}
c=d+48\left( \frac{Q}{\gamma }-\frac{a}{\gamma ^{2}}\right)  \label{c}
\end{equation}
which does not include the ghost contribution $c_{ghost}=-26$.

On the other hand, the interaction term $S_{I}$ is not an integral of
primary fields as it occurs in the usual treatment of linear dilaton CFTs. In fact, the operator 
\begin{equation}
\mathcal{T}_{\Lambda }(z)=\left( \xi \Lambda X^{0}(z)+\Lambda X^{1}(z)+2\mu
\right) e^{\kappa X^{0}(z)+\kappa X^{1}(z)}  \label{taq}
\end{equation}
turns out to be a logarithmic operator as it can be shown by the presence
of non-diagonal terms in the operator product expansion (OPE) with the
stress-tensor of the free field theory; namely 
\begin{equation}
T(z)\mathcal{T}_{\Lambda }(w)=h\frac{\mathcal{T}_{\Lambda }(w)}{(z-w)^{2}}%
+\varepsilon \frac{\mathcal{T}_{\Lambda =0}(w)}{(z-w)^{2}}+\frac{\partial 
\mathcal{T}_{\Lambda }(w)}{(z-w)}+...  \label{ope1}
\end{equation}
\begin{equation}
T(z)\mathcal{T}_{\Lambda =0}(w)=h\frac{\mathcal{T}_{\Lambda =0}(w)}{(z-w)^{2}%
}+\frac{\partial \mathcal{T}_{\Lambda =0}(w)}{(z-w)}+...  \label{ope2}
\end{equation}
where
\begin{equation}
\mathcal{T}_{\Lambda =0}(z)=2\mu e^{\kappa (X^{0}+X^{1})} , \label{La14bis}
\end{equation}
and where 
\begin{eqnarray}
h &=&-\frac{\kappa }{2}\left( Q_{0}-Q_{1}\right)  \label{peso} \\
\varepsilon &=&\frac{\Lambda }{2\mu }\left( \kappa \left( \xi -1\right) -%
\frac{\xi }{2}Q_{0}+\frac{1}{2}Q_{1}\right)  .\label{log}
\end{eqnarray}

Unlike (\ref{taq}), (\ref{La14bis})which is actually a primary field with conformal dimension $h$. For (\ref{taq}) to be a primary operator we should, in addition, to impose $\epsilon =0$. Besides, in order to realize the interacting theory as a 2D CFT describing a non-critical
string theory, we have to impose the marginal condition $h=1$, which is directly
satisfied by the original definition of the Liouville cosmological term,
the latter being proportional to $e^{\kappa (X^{0}+X^{1})}$. Here, the question might arise as to whether
the constraints $h=1$ and $\varepsilon =0$ actually represent additional restrictions
that need to be imposed on the theory. This is actually the case. These constraints
seem to be necessary conditions for the model
to yield a consistent non-critical theory. Our
analysis of the $\beta $-function equations, which we will carry out in the next section, shows how conditions (\ref{peso}) and (\ref{log}) are reobtained as necessary conditions conformal invariance.

As said above, in order to render $\mathcal{T}_{\Lambda }(w)$ a
primary operator, we also need to impose the constraint $\varepsilon =0$ which
turns the operator (\ref{taq}) into a diagonal (prelogarithmic) operator.
Thus, we obtain the following condition 
\begin{equation}
2\Lambda \kappa \left( \xi -1\right) =\Lambda \left( \xi Q_{0}-Q_{1}\right)
\end{equation}
which can be written in terms of the original parameters as follows 
\begin{equation}
\tilde{\Lambda}\left( a+b\gamma \right) \left( \gamma ^{2}-2Q\gamma
+4a\right) =0  \label{a}
\end{equation}
This equation, besides the trivial solution $\Lambda =0$, has solutions of the
form 
\begin{equation}
\gamma =Q\pm \sqrt{Q^{2}-4a}  \label{result}
\end{equation}
This is an additional restriction on the value of central charge. It permits to consider the interaction term in the context of the CFT
method. Notice that $\xi =1$ also solves the equation (\ref{a}); however, it
is not an allowed value since in such case the transformation (\ref{var1})-(%
\ref{var2}) becomes singular and the equivalence between the original
gravity action and the linear dilaton background (\ref{sx}) breaks down.

The introduction of the constraint (\ref{result}) is not a minor detail;
notice that, when replaced in (\ref{c}), the central charge of the model
becomes 
\begin{equation}
c=d+24+\frac{48a}{\gamma ^{2}}  \label{cc}
\end{equation}
which can not be set to $26$ for the case $d\geq 2$ provided $a>0$. On the other
hand, this constraint excludes particular cases like $Q=a$, $\gamma =2$,
which were considered in the $\Lambda =0$ case as examples of a critical
string theory \cite{mazzitelli}. The $Q=\frac{3}{2}$ model
receives particular interest in the context of studies of $W_{3}$ gravity 
\cite{mohammedi}. 

Notice that condition (\ref{result}) is not present in the 
$\Lambda =0$ case since the interaction term becomes primary at that point of the moduli space.
Of course, (\ref{result}) can be obtained directly by using the original
variables, for instance by solving the linearized tachyon equation in the
case of non-vanishing $\Lambda $ in the coordinates adopted in reference 
\cite{mazzitelli}. In terms of the nomenclature of reference 
\cite{chamseddine}, however, this condition cannot be satisfied because of the
particular fixation of the value of the background charge $Q$.

Then, we are led to state that an additional condition
on the central charge arises whenever a non-vanishing cosmological constant is considered in the Jackiw-Teitelboim model coupled to Liouville CFT. This new
restriction is independent of the explicit value of $\Lambda $.

This proposal needs to be contrasted with the \textit{healing} proposed in
reference \cite{chamseddine2}, where, instead imposing condition (\ref
{result}), it was suggested to change the tachyon field by adding a new term
proportional to $\sigma e^{\gamma \sigma }$, which could appear in certain
unknown step in the renormalization procedure applied to the non-local
Polyakov action.

It would be important to mention that similar operators $\sim \varphi
e^{\gamma \sigma }$ appear in other treatments of two-dimensional gravity;
for instance, see \cite{andreev}.

\section{Linearized tachyon field equation}

Alternatively, we could analyze the above restriction on the central
charge by computing the $\beta $-function equations of the model.
In fact, as early observed in \cite{chamseddine,mazzitelli}, the whole
action $S_{0}+S_{I}$ can be interpreted as a worldsheet string action
formulated on Minkowski target spacetime in presence of non-trivial
background fields $\Phi (X)$ (the dilaton) and $\mathcal{T}(X)$ (the
tachyon), whose classical configurations are given by 
\begin{eqnarray}
\Phi (X) &=&\frac{\lambda }{2}+\frac{1}{2}Q_{\mu }X^{\mu }  \label{solu} \\
\mathcal{T}_{\Lambda }(X) &=&\left( \Lambda \zeta _{\mu }X^{\mu }+2\mu
\right) e^{\kappa _{\nu }X^{\nu }}  \label{solution}
\end{eqnarray}
Notice that the functional form (\ref{solution}) is not a particular case of the theory of gravity defined by action $S_{eff}$;
this form is rather a more general \textit{ansatz} that includes the
particular case of the tachyon configuration appearing in (\ref{interaction}%
).

The $\beta $-function equations $\beta ^{\Phi }=\beta _{\mu \nu }^{G}=\beta
_{\mu \nu }^{B}=0$ are immediately satisfied at the lowest order in $\alpha
^{\prime }$ by the condition $c=26$. However, the linearized tachyon field
equation $\beta ^{\mathcal{T}}=0$ demands additional
functional relations between the parameters; namely 
\begin{equation}
\beta ^{\mathcal{T}}=-\frac{1}{2}\nabla _{\mu }\nabla ^{\mu }\mathcal{T}%
_{\Lambda }+\nabla _{\mu }\Phi \nabla ^{\mu }\mathcal{T}_{\Lambda }-\frac{2}{%
\alpha ^{\prime }}\mathcal{T}_{\Lambda }+...=0  \label{beta}
\end{equation}
where we neglected highest orders in the $\alpha ^{\prime }$ expansion. Therefore, we can adopt the convention $\alpha ^{\prime }=2$. (Notice that our
convention will be consistent with $\alpha ^{\prime }=2$. For instance, see
Eqs. (2.5.1a), (3.7.6), and (9.9.2) of Ref. \cite{polchinski}. Similarly
to the convention adopted in \cite{mazzitelli}, our notation is consistent
with the one used in Ref. \cite{tseytlin}.)

The configuration (\ref{solution}) represents a small tachyon, since we are
considering perturbation theory in $\alpha ^{\prime }$. The leading terms in
the tachyon effective action are linear in $\mathcal{T}_{\Lambda }$;
neglecting the higher powers of the tachyon field is consistent with dealing
with such field as a perturbation of the background fields that source
of geometry. Therefore, it is sufficient to solve the linearized equation 
\begin{equation}
-\partial _{\mu }\partial ^{\mu }\mathcal{T}_{\Lambda }+Q_{\mu }\partial
^{\mu }\mathcal{T}_{\Lambda }-2\mathcal{T}_{\Lambda }=0
\end{equation}
which admits a solution of the form (\ref{solution}) if the following
constraints hold 
\begin{eqnarray}
\kappa _{\mu }(Q^{\mu }-\kappa ^{\mu }) &=&2  \label{w1} \\
\zeta _{\mu }\left( Q^{\mu }-2\kappa ^{\mu }\right) &=&0  . \label{w2}
\end{eqnarray}

From (\ref{w1}), we can see that in the absence of the linear
dilaton field ($Q_{\mu }=0$) the mass-shell condition for the tachyon field $%
\kappa _{\mu }\kappa ^{\mu }=-2$ is recovered. Here, we have $%
c_{gravity}=2+3Q_{\mu }Q^{\mu }$.

In our case, we are dealing with the particular configuration
\begin{eqnarray*}
\zeta _{\mu }\zeta ^{\mu } &=&\Lambda ^{2}\frac{4a}{b\gamma }\left( \frac{4a%
}{b\gamma }+1\right) \\
Q_{\mu }Q^{\mu } &=&\frac{16}{\gamma }\left( Q-\frac{a}{\gamma }\right) \\
\kappa _{\mu }\kappa ^{\mu } &=&0
\end{eqnarray*}
$i.e.$ in the coordinates defined above we have $\zeta _{\mu }=\Lambda (\xi
,1,0,0...0)$, with the background charge $Q_{\mu }=(Q_{0},Q_{1},0,0,...0)$, and the light-like momentum $\kappa ^{\mu }=(-\kappa
,\kappa ,0,0,...0).$ Then, we can notice that, in terms of the original
parameters, the constraint (\ref{w2}) demands 
\begin{equation}
\tilde{\Lambda}\left( a+\gamma b\right) \left( \gamma ^{2}-2Q\gamma
+4a\right) =0,  \label{tuco}
\end{equation}
which exactly agrees with the condition (\ref{a}). In other words, the $\beta$-function equation coincides with the condition we had obtained in the previous section by demanding the prelogarithmic condition $\varepsilon =0$ for
the interaction term in the CFT action. Of course, the agreement between (%
\ref{tuco}) and (\ref{a}) should not surprise us since the linearized
tachyon field equation manifestly represents the requirements of conformal
invariance for the interaction term \cite{tseytlin,becker,witten}. Rather, this should be regarded as a consistency check. 

Therefore, we find that the $\beta $-function equation for the tachyon field, $\beta ^{\mathcal{T}}=0$, allows to obtain the prelogarithmic condition $\varepsilon =0$ for the operator $\mathcal{T}_{\Lambda }$ in the worldsheet
CFT. The logarithmic part of the OPE $T(z)\mathcal{T}_{\Lambda }(w)$ is parametrized by the
values ($Q,\gamma $), and these diagonalize the $L_{0}$ operator in the
critical point satisfying (\ref{result}). We emphasize that this argumentation is
independent of the explicit value of the non-vanishing cosmological constant 
$\Lambda $.

Summarizing, a new restriction appears whenever a non-vanishing cosmological constant is
considered in the model. In the general case, it is not possible to interpret the theory
as a critical string theory as it was done in the $\Lambda =0$\ case, which
appears now as a very special case.

As in the case of the exponential functional form of $\mathcal{T}_{\Lambda
=0}$, the above solution for the tachyon field in the linear dilaton
background solves the lowest orders in $\alpha ^{\prime }$ of $\beta $%
-function equations only for the lowest order in powers of the field $%
\mathcal{T}_{\Lambda }$.

The solution (\ref{solution}) is a simple generalization of the
David-Distler-Kawai background \cite{david,distler}. From (\ref{c}), (\ref
{w1}) and (\ref{w2}), we can write
\begin{equation}
\kappa _{\mu }\kappa ^{\mu }-\kappa _{\mu }Q^{\mu }+\frac{1}{4}Q_{\mu
}Q^{\mu }+\frac{d-2}{12}=0
\end{equation}
With the intention to analyze the general aspects of the solutions of such
equation of motion we can see that in the case of $\kappa ^{\mu }$ being a
time-like vector of the particular form $\kappa ^{\mu }=(\kappa ,0,0,...0)$,
the solution for the linearized tachyon field takes the form $\kappa ^{\mu
}=-\frac{1}{2}( Q_{0}\pm \sqrt{Q_{1}^{2}+\frac{d-2}{6}}) \delta
^{0\mu },$ whose qualitative behavior depends on the relative value of $d$
and $Q_{1\text{.}}$ On the other hand, in the case of $\kappa ^{\mu }$ being a
space-like vector, $v.g.$ the case $\kappa ^{\mu }=(0,\kappa ,0,...0)$, we
have $\kappa ^{\mu }=\frac{1}{2}( Q_{1}\pm \sqrt{Q_{0}^{2}-\frac{d-2}{6}%
}) \delta ^{1\mu }.$ The physical aspects of such solutions were
extensively studied within the context of non-critical string theory where
it was discussed how the tachyon field acts as a barrier of potential
competing with the action of the linear dilaton background, see for
instance \cite{david,polchinski,distler,becker,witten}.

Moreover, the case we are interested on is the analogous to the
Jackiw-Teitelboim model coupled to Liouville theory, which corresponds to
the light-like case $\kappa _{\mu }\kappa ^{\mu }=0$. In this third case, the
solutions fall into the region described by the constraint 
\begin{equation}
\kappa _{\mu }Q^{\mu }=2  \label{n29}
\end{equation}
which is precisely the mass shell condition $h=1$ for the tachyon field in
this particular dilatonic configuration. Here, the null condition $\kappa _{\mu }\kappa
^{\mu }=0$ appears as the manifestation of the neutralization effect of
Liouville mode induced by the constant curvature constraint, which was
studied in reference \cite{chamseddine} within the context of the path
integral approach. Light-like condition (\ref{n29}) can be
seen as a change in the propagator of Liouville mode \cite
{chamseddine,mazzitelli,david} and it leads to the linear functional form
for the conformal dimension of the operators with the form $\sim e^{\kappa
(X^{0}+X^{1})}$, instead the usual quadratic dependence that leads to the
definition of two-different screening charges, Weyl-like reflection relations
and the existence of conjugated pictures in the case of similar CFTs. Let us postpone some comments about this for the next section. Here, the intention was merely exploring the general features of including a
non-vanishing cosmological constant $\Lambda $ in the context of the stringy
interpretation of the Jackiw-Teitelboim gravity coupled to Liouville theory.

\section{Three-point correlation functions}

Now, by using this CFT description of the model, it is feasible to follow
the steps of the Coulomb gas prescription and write down
explicit expressions for the integral representation of correlation
functions in the theory defined by action (\ref{sx}). For instance, in
reference \cite{chamseddine} it was pointed out the importance of extending
the CFT analysis with the goal of computing correlation functions in
this model. On the other hand, in \cite{dotsenko}, a similar treatment of
three-point functions was presented for the case of minimal models coupled
to two-dimensional gravity. Indeed, thanks to the formulation of the theory in the 2D CFT language, it is relatively easy to compute the
three-point functions in the Jackiw-Teitelboim gravity coupled to Liouville
theory for the case $\Lambda =0$ by using the free field realization.
In fact, the $N$-point correlation functions of primary fields $\Phi
_{k}(z)=\exp ik_{\mu }X^{\mu }(z)$ on the sphere are described by the
following form 
\begin{equation*}
\left\langle \prod_{i=1}^{N}\Phi _{\alpha ^{(i)},\beta
^{(i)},k^{(i)}}(z_{i})\right\rangle _{S_{eff}}=e^{-2\lambda }\mu ^{t}\Gamma
(-t)\prod_{r=1}^{t}\int d^{2}w_{r}\left\langle \prod_{i=1}^{N}\Phi _{\alpha
^{(i)},\beta ^{(i)},k^{(i)}}(z_{i})\prod_{r=1}^{t}\mathcal{T}_{\Lambda
=0}(w_{r})\right\rangle _{S_{0}}
\end{equation*}
where standard formulae of the path integral realization has been considered (see,
for instance \cite{becker,goulian,francesco,dotsenko2,dotsenko}) and we have
defined $\alpha =ik_{0}$, $\beta =ik_{1}$ . By using the projective
invariance, three different inserting points of vertex operators can be
fixed on the worldsheet in order to cancel the volume of the Killing conformal group, $SL(2,C)$; $v.g.$ we can set $\left( z_{1},z_{2},z_{3}\right) =\left( 0,1,\infty
\right)$ as usual. Then, we can write the $N$-point correlation
functions $G^{N}(z_{1},z_{2},...z_{N})$ as 
\begin{equation*}
G_{k_{\mu _{1}}^{(1)},k_{\mu _{2}}^{(2)},...k_{\mu
_{N}}^{(N)}}^{N}(z_{1},z_{2},...z_{N})=\left\langle \prod_{i=1}^{N}\Phi
_{\alpha ^{(i)},\beta ^{(i)},k^{(i)}}(z_{i})\right\rangle _{S_{eff}}=
\end{equation*}
\begin{eqnarray*}
&=&\frac{(2\pi )^{d-1}(2\mu )^{t}e^{-2\lambda }}{\kappa }\Gamma
(-t)\prod_{i>j}^{N,N-1}\left| z_{i}-z_{j}\right| ^{2\left(
k^{(i)}k^{(j)}+\alpha ^{(i)}\alpha ^{(j)}-\beta ^{(i)}\beta ^{(j)}\right)
}\times \\
&&\times \prod_{r=1}^{t}\int d^{2}w_{r}\prod_{r=1}^{t}\prod_{i=1}^{N}\left|
z_{i}-w_{r}\right| ^{2\kappa (\alpha ^{(i)}-\beta ^{(i)})}\times \\
&&\delta \left( \sum_{i=1}^{N}(\alpha ^{(i)}-\beta
^{(i)})+Q_{0}-Q_{1}\right) \delta ^{(d-2)}\left( \sum_{i=1}^{N}k^{(i)}\right)
\end{eqnarray*}
where we denoted 
\begin{equation}
t=\frac{1}{2\kappa }\left(
Q_{0}+Q_{1}-\sum_{i=1}^{N}(\alpha ^{(i)}+\beta ^{(i)})\right) . 
\end{equation}
In the expression above, the delta
functions and the global factor ${(2\pi )^{d-1}(2\mu )^{t}}/{\kappa }$
appear through the integration over the zero-modes of $X^{\mu }$ fields, while
the factor $e^{-2\lambda }$ appears because of the Einstein-Hilbert term, which gives such exponent as the genus-zero contribution in the functional measure $e^{-S_{eff}}$. In order the correlation functions not to vanish, the condition 
\begin{equation}\sum_{i=1}^{N}\left( \alpha
^{(i)}-\beta ^{(i)}\right) =Q_{0}-Q_{1}
\end{equation} 
is required. Therefore, we can write the
conservation laws as 
\begin{equation}
\sum_{i=1}^{N}\alpha ^{(i)}\pm \sum_{i=1}^{N}\beta ^{(i)}=\left(
Q_{0}-t\kappa \right) \pm (Q_{1}-t\kappa )  \label{numeralon}
\end{equation}
with $t$ being a non-negative integer.

Notice that mixed terms of the form $\left| w_{r}-w_{r^{\prime }}\right|
^{2\delta }$ do not exist because of the particular configuration $\delta
=-\kappa _{\mu }\kappa ^{\mu }=0$. This makes the integration over
the different screening insertions to factorize, yielding a direct product of independent integrals. This is a manifestation of
the neutralization effect of the Liouville mode pointed out before; $i.e.$
in the case of three-point functions, due to the light-like condition $\kappa _{\mu }\kappa
^{\mu }=0$, the Dotsenko-Fateev type integrals 
\cite{goulian,francesco,dotsenko2,dotsenko} become a product of
Virasoro-Shapiro integrals. Then, we can explicitly compute the three-point correlators as follows 
\begin{equation*}
G_{k_{\mu _{1}}^{(1)},k_{\mu _{2}}^{(2)},k_{\mu _{3}}^{(3)}}^{3}(0,1,\infty
)=\left\langle \Phi _{\alpha ^{(1)},\beta ^{(1)},k^{(1)}}(0)\Phi _{\alpha
^{(2)},\beta ^{(2)},k^{(2)}}(1)\Phi _{\alpha ^{(3)},\beta
^{(3)},k^{(3)}}(\infty )\right\rangle =
\end{equation*}
\begin{eqnarray*}
&=&\frac{(2\pi )^{d-1}\left( 2\mu \right) ^{t}e^{-2\lambda }}{\kappa }\Gamma
(-t)\prod_{r=1}^{t}\int d^{2}w_{r}\prod_{r=1}^{t}\left| w_{r}\right|
^{2\kappa (\alpha ^{(1)}-\beta ^{(1)})}\left| 1-w_{r}\right| ^{2\kappa
(\alpha ^{(2)}-\beta ^{(2)})}\times \\
&&\times \delta \left( \sum_{i=1}^{3}(\alpha ^{(i)}-\beta
^{(i)})+Q_{0}-Q_{1}\right) \delta ^{(d-2)}\left( \sum_{i=1}^{3}k^{(i)}\right)
\end{eqnarray*}
obtaining
\begin{equation*}
G_{k_{\mu _{1}}^{(1)},k_{\mu _{2}}^{(2)},k_{\mu _{3}}^{(3)}}^{3}=\frac{(2\pi
)^{d-1}e^{-2\lambda }}{\kappa }\Gamma \left( \frac{1}{\kappa }\left( \alpha
^{(1)}+\alpha ^{(2)}+\alpha ^{(3)}-Q_{0}\right) \right) \times
\end{equation*}
\begin{equation*}
\times \left( 2\pi \mu \frac{\gamma (1+\kappa (\alpha ^{(1)}-\beta
^{(1)}))\gamma (1+\kappa (\alpha ^{(2)}-\beta ^{(2)}))}{\gamma (2+\kappa
(\alpha ^{(1)}+\alpha ^{(2)}-\beta ^{(1)}-\beta ^{(2)}))}\right) ^{\frac{1}{%
\kappa }\left( Q_{0}-\sum_{i=1}^{3}\alpha ^{(i)}\right) }
\end{equation*}
where $\gamma (x)={\Gamma (x)}/{\Gamma (1-x)}$. By considering the
condition $Q_{1}-Q_{0}={2}/{\kappa }$ and noting that $\gamma
^{-1}(x)=\gamma (1-x),$ we can write 
\begin{equation}
\prod_{i=1}^{3}\gamma ^{-1}(\kappa
(\beta ^{(i)}-\alpha ^{(i)}))=\frac{1}{2}\prod_{i\neq j}\gamma (\kappa
(\beta ^{(i)}+\beta ^{(j)}-\alpha ^{(i)}-\alpha ^{(j)})-1);
\end{equation}
then, the above
expression can be rewritten as follows
\begin{equation}
G_{k_{\mu _{1}}^{(1)},k_{\mu _{2}}^{(2)},k_{\mu _{3}}^{(3)}}^{3}=\frac{(2\pi
)^{d-1}e^{-2\lambda }}{\kappa }\Gamma \left( \sum_{i=1}^{3}\frac{\alpha
^{(i)}}{\kappa }-\frac{Q_{0}}{\kappa }\right) \left( 2\pi \mu \prod_{i=1}^{3}%
\frac{\Gamma \left( 1+\kappa (\alpha ^{(i)}-\beta ^{(i)})\right) }{\Gamma
\left( \kappa (\beta ^{(i)}-\alpha ^{(i)})\right) }\right) ^{p }  \label{yupi}
\end{equation}
with $p=( Q_{0}-\sum_{i=1}^{3}\alpha ^{(i)})/{\kappa 
}$, satisfying (\ref{numeralon}). This is the general form for
three-point correlation functions which has been computed by using the Coulomb gas prescription.

Before concluding, let us make a last remark about the role of the field $\varphi $ in the story. To this end, let us consider an additional dynamical term for such field (see also \cite{mazzitelli}); namely 
\begin{equation}
S_{\varphi }=\frac{g}{4\pi }\int d^{2}z\sqrt{h}h^{\alpha \beta }\partial
_{\alpha }\varphi \partial _{\beta }\varphi . \label{din}
\end{equation}

If such a term is included, the case $\xi \rightarrow \xi =-%
\frac{2\gamma a+b}{2g+b}$ needs to be considered in (\ref{var2}) in order to
diagonalize the metric of target spacetime. In this case, we would obtain a
non-vanishing correlator $\left\langle \sigma (z)\sigma (w)\right\rangle $
for the Liouville mode ($i.e.$ $\kappa _{\mu }\kappa ^{\mu }\neq 0$). Then,
the inclusion of a dynamical term for $\varphi $ field restores the
influence of the Liouville field $\sigma $, which had became harmless due to the constant curvature constraint. Besides, if (\ref{din})
is included, the three-point correlators are given in terms of
Dotsenko-Fateev type integrals 
\begin{equation}
\prod_{r=1}^{t}\int
d^{2}w_{r}\prod_{r=1}^{t}\left| w_{r}\right| ^{2(\kappa _{0}\alpha
^{(1)}-\kappa _{1}\beta ^{(1)})}\left| 1-w_{r}\right| ^{2(\kappa _{0}\alpha
^{(2)}-\kappa _{1}\beta ^{(2)})}\prod_{r^{\prime }<r}^{t-1,t}\left|
w_{r^{\prime }}-w_{r}\right| ^{2(\kappa _{0}^{2}-\kappa _{1}^{2})} ,
\end{equation}
which
can be performed by using the result of reference \cite{dotsenko2}
(specifically, see Eq. (B.9) therein). In such case, a global factor of
the form 
\begin{equation}
\Gamma (t+1)\prod_{r=0}^{t}\frac{\gamma (r(\kappa _{0}^{2}-\kappa
_{1}^{2})/2)}{\gamma ((\kappa _{0}^{2}-\kappa _{1}^{2})/2)} 
\end{equation}
stands in the
expression for three-point correlators; it is easy to see that in the $%
\kappa _{\mu }\kappa ^{\mu }\rightarrow 0$ limit, where the Liouville mode
becomes harmless again, the expression (\ref{yupi}) is recovered.

\section{Conclusions}

In this paper, the implications of considering a non-vanishing cosmological constant in
Jackiw-Teitelboim two-dimensional theory of gravity coupled to Liouville
theory have been discussed. This was done by treating the model
as a non-critical string theory. Within the context of this stringy
interpretation, we emphasize that a non-zero value for $\Lambda $ leads to revise the analysis of the parameter space and coupling of both actions. Additional restrictions on the value of the central charge of the
model need to be considered even if the conformal anomaly turns out to be independent
of the explicit value of $\Lambda $. The constraint equations that realize the
additional restrictions agree with the necessary conditions for the
tachyon field $\mathcal{T}_{\Lambda }(z)$ to be a primary (prelogarithmic) marginal operator \cite{kogan} of the worldsheet
CFT. We studied the $\beta$-function equation and observes that the linearized field equation for the tachyon field leads to impose the same diagonal condition for the operator product expansion $T(z)\mathcal{T}_{\Lambda }(w)$.

We have seen how the coupling between two-dimensional gravity and Liouville
theory leads to the neutralization of the Liouville mode. This had been pointed
out in reference \cite{chamseddine}, where it was explained how the constant
curvature constraint renders the Liouville action \textit{harmless} by 
\textit{trivializating the} $\sigma $ \textit{dependence}. This
neutralization effect is reflected, within the context of the conformal
field theory description, by means of the light-like condition $\kappa _{\mu
}\kappa ^{\mu }=0$, which permits to realize the correlation functions
in terms of a direct product of multiple integrals over the whole complex
plane. Then, explicit expression for three-point functions were obtained in
terms of Virasoro-Shapiro integrals, which emerge from the neutralization
condition imposed on the Dotsenko-Fateev type realization. These results
have been obtained by means of the description of the model in the context
of the Coulomb gas formalism, which, once again, turns out to be a fruitful
method to study several features of two-dimensional quantum gravity.

\bigskip \vskip0.5 cm

I wish to thank Noureddine Mohammedi and Mariano Galvagno.

\end{document}